\documentstyle[version2,preprint,aps]{revtex} 
                                                                        
\begin{document}                                                        
                                                                        
\draft                                                                  

\begin{title}
Spin bags in the doped $t$$-$$J$ model
\end{title}

\author{R. Eder and Y. Ohta}

\begin{instit}
Department of Applied Physics, Nagoya University, Nagoya 464-01, Japan
\end{instit}

\begin{abstract}
We present a nonperturbative method for deriving
a quasiparticle description of the low-energy excitations in
the $t$$-$$J$ model for strongly correlated electrons.
Using the exact diagonalization technique
we evaluated exactly the
spectral functions of composite operators, which
describe an electron or hole dressed by antiferromagnetic
spin fluctuations as expected in the string or spin bag picture. 
For hole doping up to $1/8$,
use of the composite operators leads to a 
drastic simplification of the single particle spectral function:
at half-filling it takes free-particle form,
for the doped case it resembles a system of weakly
interacting Fermions corresponding to the doped holes.
We conclude that for all doping levels under study,
the elementary electronic excitations next to the Fermi level
are adequately described by the antiferromagnetic 
spin fluctuation picture.
\end{abstract} 
                                                                               

Despite great efforts
the unusual properties of high-temperature superconductors
remain a largely unresolved problem. 
There is general agreement that the electrons in these materials
are strongly correlated, so that marked deviations from
the single particle picture are to be expected.
An important step in
setting up a successfull theory of both, their normal
and superconducting state, therefore would be to find
a description in terms of elementary excitations i.e.
weakly or moderately interacting quasiparticles.
We present an exact diagonalization study
of the $t$$-$$J$ model
which is specifically aimed at finding such a description.
We construct composite operators which reduce or increase
the electron number by one and simultaneously
rearrange the spins in the
neighborhood of the newly created hole/electron so as to
simulate the `cloud of spin defects' surrounding the hole.
For doping levels up to $1/8$, the spectral function
of the composite operators then takes the
form expected for weakly interacting Fermions,
with a well defined `band' right at the Fermi level.
The $t$$-$$J$ model reads:
\[ H =                                                    
 -t \sum_{< i,j >, \sigma}                                         
( \hat{c}_{i, \sigma}^\dagger \hat{c}_{j, \sigma}  +  H.c. )
 + J \sum_{< i,j >} [\;\bbox{S}_i \cdot
 \bbox{S}_j
 - \frac{n_i n_j}{4}\;].
\]
The $\bbox{S_i}$ are the electronic                                        
spin operators and                                                             
the sum over $<i,j>$ stands for a summation                                    
over all pairs of nearest neighbors on a                                       
two dimensional square lattice.                                                
The operators $\hat{c}^\dagger_{i,\sigma}$
are expressed in terms of ordinary fermion                                  
operators as $c^\dagger_{i,\sigma}(1-n_{i,-\sigma})$.
We present results for a $4$$\times$$4$ cluster of this model
with $t/J$$=$$4$, similar results have been obtained also for 
different values of $t/J$.\\
A single hole in the half-filled band can be well described by the
string\cite{Shr1,Trugman,Maekawa} 
or spin bag\cite{Schriefferetal} picture, where the 
hole is dressed by antiferromagnetic spin fluctuations.
With this in mind, we make the following ansatz for a
`spin bag operator'
($N(j)$ denotes the nearest neighbors of $j$): 
\begin{eqnarray}
\tilde{c}_{\bbox{k},\uparrow} &=& 
\frac{1}{\sqrt{N}}
\sum_{j}  \sum_{\lambda=0}^{\lambda_{max}}
e^{i \bbox{k} \cdot \bbox{R}_j} \alpha_\lambda(\bbox{k})
\;A_{j,\lambda},\nonumber \\
A_{j,0} &=& 
\hat{c}_{j,\uparrow}, \nonumber \\
A_{j,1} &=& \sum_{k\in N(j)} S_j^- 
\hat{c}_{k,\downarrow}, 
\nonumber \\
A_{j,2} &=& \sum_{k\in N(j)} \sum_{l\in N(k)} 
S_j^- S_k^+ \hat{c}_{l,\uparrow}.
\nonumber \\
A_{j,3} &=& \sum_{k\in N(j)} \sum_{l\in N(k)} \sum_{m\in N(l)} 
S_j^- S_k^+ S_l^- \hat{c}_{m,\downarrow}.
\label{QP}
\end{eqnarray}
When acting on the N\'eel state, the operator
$A_{j,\lambda}$  creates all strings of length $\lambda$ which begin
at site $j$, $\tilde{c}_{\bbox{k},\sigma}$ reproduces a 
simple trial wave function for a single hole\cite{I}.
$A_{j,\lambda}$ also can be thought of as
having been generated by $\lambda$-fold
commutation of $\hat{c}_{j,\uparrow}$ with the
kinetic energy, a procedure suggested by 
Dagotto and Schrieffer\cite{DagottoSchrieffer}. 
The parameters $\alpha_\lambda(\bbox{k})$ are 
determined\cite{DagottoSchrieffer}
from the requirement
that the state $\tilde{c}_{\bbox{k},\sigma}|\Psi_0^{(0h)}\rangle$
(where $|\Psi_0^{(0h)}\rangle$ denotes the 
half-filled ground state) 
has norm $1$ and maximum overlap with the lowest
totally symmetric single-hole eigenstate
with momentum $\bbox{k}$, $|\Psi_0^{(1h)}(\bbox{k})\rangle$.
If we denote the Fourier transform of $A_{j,\nu}$ by
$A_{\bbox{k},\nu}$ and introduce
\begin{eqnarray}
n_\nu &=&
\langle \Psi_0^{(1h)}(\bbox{k})|A_{\bbox{k},\nu} |\Psi_0^{(0h)}\rangle,
\nonumber \\
A_{\mu,\nu} &=& n_\mu^* n_\nu,
\nonumber \\
B_{\mu,\nu} &=&
\langle \Psi_0^{(0h)}| A_{\bbox{k},\mu}^\dagger A_{\bbox{k},\nu}
|\Psi_0^{(0h)}\rangle,
\end{eqnarray}
the $\alpha$'s can be obtained by solution of the generalized
eigenvalue problem $A x = \lambda B x$.
We use  $\lambda_{max}$$=$$3$, so that the
$\tilde{c}_{\bbox{k},\sigma}$ effectively contain 
$3$ free parameters.
Having fixed the $\alpha$'s, we compute the spectral function
\[
A_- (\bbox{k}, -\omega) =
\sum_{\nu} 
| \langle \Psi_\nu^{(1h)} | \tilde{c}_{\bbox{k},\sigma} |
\Psi_0^{(0h)} \rangle |^2
\delta( \omega - ( E_\nu^{(1h)} - E_0^{(0h)} )),
\]
via the standard Lanczos method.
Here $|\Psi_\nu^{(nh)} \rangle $ denotes
the $\nu^{th}$ eigenstate with
$n$ holes and  $E_\nu^{(nh)}$ the corresponding energy 
(in particular $\nu=0$ implies the ground state).
This spectral function
is shown in Fig. \ref{spec0} 
and compared to that of the `string-0 operator'
$\sqrt{2} \hat{c}_{\bbox{k},\sigma}$,
which up to a factor $2$ equals the
usual photoemission spectrum.
Despite the small number of adjustable parameters,
use of the spin bag operators
brings the spectral function
to almost perfect free-particle form:
the incoherent continua present in the
bare electron spectra are removed, the spectral 
weight being concentrated essentially in one sharp peak
with a well-defined next-nearest neighbor dispersion.
This can be understood if one
attributes the incoherent continua 
to the retracable-path-type
motion\cite{BrinkmanRice} of the hole. Since the
$A_{j,\mu}$ describe precisely the forward and backward 
hopping of the hole along a track of N\'eel ordered spins,
this type of hole-motion
is already `incorporated' into the definition of 
the $\tilde{c}_{\bbox{k},\sigma}$, so that their spectra
emphasize the coherent hole motion. 
Some care is necessary:
for momentum $(\pi,\pi)$ the lowest totally symmetric single-hole
eigenstate has spin $\frac{3}{2}$, so that it can not be 
observed in the spectrum of the bare electron operator.
On the other hand, since the $\tilde{c}_{\vec{k},\sigma}$
are not vector operators under spin rotations, it can be observed in 
their spectrum. Since this state fits very well into the 
next-nearest neighbor hopping
dispersion relation expected for a single hole, we believe that
the fact that it has spin $3/2$ in the $4$$\times$$4$ cluster
is a finite size effect. We do not have any proof for that,
so results for $(\pi,\pi)$ should be considered with
care; however, none of the conclusions to be presented below
depends crucially on the form of the spectra for this momentum.\\ 
So far, we have merely demonstrated the quality
of the string picture at half-filling which may not be very
surprising; a much more
interesting question is, whether this
description of the states next to the Fermi 
level remains valid upon doping.
Exact diagonalization offers a very direct and natural way
to check this ssue, namely to
evaluate the spectra of the $\tilde{c}_{\vec{k},\sigma}$
for a doped rather than the half-filled ground state.
For the $\alpha_\lambda(\bbox{k})$'s 
we thereby retain the values optimized at half-filling
(it would be easy to recalculate the
$\alpha$'s for the doped ground state but an important question 
is whether there is some continuity in the development of the
electronic states at the Fermi level).
We begin with the ground state
with $8$ up-spin electrons,  $7$ down-spin electrons 
and momentum $(\pi/2, \pi/2)$) i.e. the single hole ground state.
In Fig. \ref{spec11} (Fig. \ref{spec12}) the spectra of
$\tilde{c}_{\vec{k},\uparrow}$
($\tilde{c}_{\vec{k},\downarrow}$)
are again compared to those of the respective string-0 operators,
$\sqrt{2} \hat{c}_{\vec{k},\uparrow}$
($\sqrt{2} \hat{c}_{\vec{k},\downarrow}$).
Quite obviously,
the $\tilde{c}_{\vec{k},\sigma}$ continue to `work':
there is the same elimination of the incoherent continua
and enhancement of the peaks at the Fermi level as for half-filling.
A novel feature is the broadening of these peaks, and
it seems natural to ascribe it to the
scattering of the added hole
from the one already present in the system.
Most important of all, however,
the spectrum for $\tilde{c}_{\vec{k},\downarrow}$
shows an unambiguous `pocket' at
$(\pi/2, \pi/2)$, the momentum of the $\downarrow$-hole
already present in the system: we clearly see the
Pauli principle working for the spin bags.
This suggest weakly interacting spin-$1/2$ Fermions 
which correspond to the doped holes
as the `effective theory' for the low-lying
states of the cluster with $2$ holes.\\
We thus push things further and proceed to
the two-hole ground state.
Having in mind the results obtained so far,
we should model it as an interacting state of two spin bags
with total momentum zero:
\begin{equation}
|\tilde{\Psi}_0^{(2h)} \rangle 
= \sum_{\bbox{k}} \Delta(\bbox{k}) 
\tilde{c}_{\bbox{k},\uparrow}
\tilde{c}_{-\bbox{k},\downarrow}
|\Psi_0^{(0h)}\rangle.
\label{GS}
\end{equation}
Using the $\tilde{c}_{\bbox{k},\sigma}$ optimized at half-filling
we consequently construct the states
$|\Phi(\bbox{k})\rangle = 
\tilde{c}_{\bbox{k},\uparrow}
\tilde{c}_{- \bbox{k},\downarrow}|\Psi_0^{(0h)}\rangle$ for all
$16$ allowed momenta in the cluster,
evaluate the matrices
$h_{\bbox{k},\bbox{k}'} = \langle \Phi(\bbox{k})| 
H |\Phi(\bbox{k}')\rangle$ and
$n_{\bbox{k},\bbox{k}'} = \langle \Phi(\bbox{k})|\
\Phi(\bbox{k}')\rangle$ and solve
the resulting eigenvalue problem  to obtain
$\Delta(\bbox{k})$. 
The estimate for the ground state energy obtained in this way
is $-8.21t$, to be compared with the exact value of $-8.81t$,
and we have
$\langle \tilde{\Psi}_0^{(2h)} |\Psi_0^{(2h)}\rangle=0.67$.
Obviously the string gound state is not really an excellent
approximation to the exact one, but it should be noted that
it has been constructed from the half-filled groundstate,
so that no relaxation of the `spin background' 
(corresponding to the collapse of long range
N\'eel order in the infinite system) is incorporated.
On the other hand, the approximate
ground state shares
some basic features of the exact one, such as
the correct $d_{x^2-y^2}$-symmetry\cite{DagottoSchrieffer}.
In Tab. \ref{nk}, the 
`spin bag momentum distribution'
$\tilde{n}(\bbox{k})=|\langle \Phi(\bbox{k})
|\tilde{\Psi}_0^{(2h)} \rangle|^2
/\langle \Phi(\bbox{k})|\Phi(\bbox{k})\rangle$,
is listed; due to the nonorthogonality of the
$|\Phi(\bbox{k})\rangle$ there exists no simple
sum rule for this quantity, so that its interpretation as
`momentum distribution' is strictly speaking questionable.
However, it can give a rough idea of the distribution 
of the spin bags in momentum space, and obviuosly
only $(\pi,0)$ and $(\pi, \pi/2)$ are appreciably
occupied\cite{PoilblancDagotto}. 
It is interesting to contrast $\tilde{n}(\bbox{k})$
with the `bare hole momentum distribution' $n_h(\bbox{k})=
\langle \hat{c}_{\vec{k},\sigma}
\hat{c}_{\vec{k},\sigma}^\dagger\rangle$ also
given in Tab. \ref{nk} both for the approximate and 
exact two-hole ground
state. Whereas the $n_h(\bbox{k})$ for the exact and approximate
ground state agree reasonably well, there is no similarity
with  $\tilde{n}(\bbox{k})$: 
$n_h(\bbox{k})$ is roughly consistent with a
free electron picture, $\tilde{n}(\bbox{k})$
would rather suggest that mainly $(\pi,0)$ is occupied by
quasiparticles.
This can be understood by recalling
that the incoherent retraceable path-type 
motion of the bare holes
(which naturally contributes to $n_h(\bbox{k})$\cite{EderBecker})
is already absorbed into the
definition of the string operators, so that
$\tilde{n}(\bbox{k})$ measures predominantly the coherent hole
motion. \\
Next, let us find an approximate
annihilation operator for the dressed holes,
i.e. an operator $\bar{c}^\dagger_{\bbox{k},\sigma}$ so that
this operator and 
$\tilde{c}_{\bbox{k},\sigma}$ mutually
undo the action of each other.
Thereby the first guess, $\bar{c}^\dagger_{\bbox{k},\sigma}$$=$$
(\tilde{c}_{\bbox{k},\sigma})^\dagger$
may not be expected to be reasonable because
the `basis set' $A_{j,\lambda}$
does not consist of operators which create electrons in
orthogonal basis functions, i.e.
$[A_{i,\mu},(A_{j,\lambda})^\dagger]_+$
$\neq$$\delta_{i,j}$$\delta_{\mu,\lambda}$.
We thus introduce a shortcut:
for $\bar{c}^\dagger_{\bbox{k},\uparrow}$ 
we make the {\em ansatz}
\[
\bar{c}^\dagger_{\bbox{k},\uparrow} 
= \frac{1}{\sqrt{N}} \sum_{j} \sum_{\lambda=0}^{\lambda_{max}} 
e^{-i \bbox{k} \cdot \bbox{R}_j} \beta_\lambda(\bbox{k})
(A_j^{(\lambda)})^\dagger,
\]
and construct the state
\[
|\Psi_c^{(2h)} \rangle = [ \bar{c}^\dagger_{\bbox{k},\sigma}, 
\tilde{c}_{\bbox{k},\sigma} ]_+
|\Psi_0^{(2h)}\rangle.
\]
If $\bar{c}^\dagger_{\bbox{k},\sigma}$ and
$\tilde{c}_{\bbox{k},\sigma}$ indeed
were adjoint fermionic creation and annihilation operators,
the anticommutator on the right hand side would be $1$ and
consequently we determine the $\beta$'s from the requirement
that $| \Psi_c^{(2h)} \rangle$ has norm $1$ and
maximum overlap with $| \Psi_0^{(2h)} \rangle$.
The actual value of the overlap then also provides a measure for
the quality of our {\em ansatz} and with 
$\lambda_{max}=3$ we indeed find
that for most $\bbox{k}$-points
$|\langle \Psi_c^{(2h)}| \Psi_0^{(2h)} \rangle|^2$$ \sim$$1$ 
(see Tab. \ref{antico}), 
so that this way of obtaining an approximation to
$\bar{c}^\dagger_{\bbox{k},\sigma}$
appears quite reasonable. Using the $\bar{c}^\dagger_{\bbox{k},\sigma}$
we now can also study the spin bag removal (electron addition)
spectra in the two-hole ground state, defined as
\[
A_+ (\bbox{k}, \omega) =
\sum_{\nu} 
| \langle \Psi_\nu^{(1h)} | \bar{c}_{\bbox{k},\sigma}^\dagger |
\Psi_0^{(2h)} \rangle |^2
\delta( \omega - ( E_\nu^{(1h)} - E_0^{(2h)} )).
\]
Thereby some care is necessary: in contrast to
the ordinary electron operators, there exists no simple
sum rule for the integrated weight of the
spin bag addition and removal spectra.
To faciliate the comparison with the usual single particle spectral
function in the following all spectra
are therefore normalized to unity.\\
Fig. \ref{spec2} then compares
the spin bag spectral function
and that of the ordinary electron operators.
The spin bag creation (electron annihilation)
spectra clearly show the 
simplification already familiar from 
the previous calculations: the incoherent continua
far from the Fermi energy are removed, the peaks near $E_f$ are
markedly enhanced. We conclude that even for this level of doping, 
the $\tilde{c}_{\bbox{k},\sigma}$
optimized at half-filling
provide a good description of the electronic
excitations closest to the Fermi energy, $E_f$.
The degree of broadening
of the spin-bag peaks is reminiscent
of a Fermi liquid:
there are sharp peaks close to $E_f$, diffuse peaks
at lower energies. For the spin bag annihilation
(electron creation) spectra use of the adjoint spin 
bag operators leads to an increase of spectral weight at $(\pi,0)$
and a marked depletion at $(\pi,\pi)$.
The resulting division of spectral weight
between spin bag addition and removal spectra
moreover is remarkably consistent with the 
`momentum distribution' $\tilde{n}(\bbox{k})$:
spin bags can be annihilated predominantly at
$\bbox{k}$$=$$(\pi,0)$ and $\bbox{k}$$=$$(\pi, \pi/2)$, 
i.e. the momenta which were most probable 
in the variational ground state $|\tilde{\Psi}_0^{(2h)} \rangle$.
All in all the doped cluster thus behaves very much like a
system of weakly interacting `effective fermions' corresponding to the
doped holes.\\
In summary we have shown that for all doping  levels studied
the use of `spin bag operators' leads to a drastic 
simplification of the spectral function. 
Since these  operators describe the modification of the
`spin background' in the immediate neighborhood of the hole,
we also
expect that they are predominantly determined by the local
spin correlations, which most probably are described adequately
by the exact diagonalization. 
It thus seems reasonable to expect that
a similar simplification of the spectral function occurs also in the
infinite system. Our nonperturbative approach
thus suggests a rather conventional
`effective theory', in which the
electronic excitations right at the Fermi level
are modelled by spin $1/2$ Fermions corresponding to the doped holes
(of course, like any other cluster calculation the present study 
cannot adress the dependence of the quasiparticle lifetime
on the distance in energy from the Fermi energy).
The remarkable degree of continuity upon doping moreover
shows that these
quasiparticles are holes dressed by antiferromagnetic
spin fluctuations in essentially the same way
as a single hole moving in an antiferromagnet, a reasonably well
understood 
problem\cite{Shr1,Trugman,Maekawa,SchmittRink,KaneLeeRead}.
The dressing of the holes with spin fluctuations 
leads to an interaction which 
favours a bound state with $d_{x^2-y^2}$ symmetry.
Our exact results thus clearly 
corroborate the basic assumptions of the
antiferromagnetic spin fluctuation theory of
high-temperature superconductivity\cite{Moriya,Monthoux}.
Another important point is, that the absorption of the
incoherent hole motion into the definition of the spin 
bag operators leads to an almost complete
`decoupling' of the quasiparticle 
spectral function and momentum distribution
from that of the bare electrons.
This suggests consequences for the interpretation of
experiments: it seems plausible that measurements 
of transport properties\cite{TrugmanI,Takagi} 
would rather probe
the `quasiparticle properties' and not resolve the
internal structure of the spin bags. On the other hand,
high-energy experiments like photoemission\cite{Olson,King}
should resolve the internal structure and
reflect properties of the bare electrons.
Discrepancies between the transport properties
and the Fermi surface measured in photoemission\cite{King}
thus may be not surprising.\\
It is a pleasure to acknowledge numerous instructive discussions with
Professor S. Maekawa. Financial support by the Japanese Society
for the Promotion of Science is most gratefully acknowledged.
\figure{Spectral function of the spin bag operators (full line)
        and ordinary electron annihilation operator (dotted line) 
        in the half-filled ground state of the $4$$\times$$4$ cluster.
\label{spec0} }
\figure{Spectral function $A_- (\bbox{k}, \omega)$
        of the spin bag operators (full line)
        and ordinary electron annihilation operator (dotted line) 
        in the single hole ground state (momentum 
        $(\pi/2,\pi/2)$) of the
        $4$$\times$$4$ cluster. The spin of the newly created hole
        is antiparallel to that of the hole already present.
\label{spec11} }
\figure{Spectral function $A_- (\bbox{k}, \omega)$ 
        of the spin bag operators (full line)
        and ordinary electron annihilation operator (dotted line) 
        in the single hole ground state (momentum 
        $(\pi/2,\pi/2)$) of the
        $4$$\times$$4$ cluster. The spin of the newly created hole
        is parallel to that of the hole already present.
\label{spec12} }
\figure{Spectral function $A_- (\bbox{k}, \omega)+
        A_+ (\bbox{k}, \omega)$ 
        of the spin bag operators (full line)
        and ordinary electron operators (dotted line) 
        in the two hole ground state
        of the $4$$\times$$4$ cluster. The Fermi energy is marked by
        a thin line and the frequency region
        $\omega$$<$$E_F$ ($\omega$$>$$E_F$) corresponds to the
        annihilation (creation) of an electron.
\label{spec2} }
\begin{table}
\caption{`Quasiparticle momentum distribution' $\tilde{n}(\bbox{k})$
in the approximate two-hole ground state and
bare hole distribution function $n_h(\bbox{k})$ for the
approximate (A) and exact (E) two-hole ground state.}
\begin{tabular}{c | c c c c c c}
$\bbox{k}$ &$(0,0)$ & $(\frac{\pi}{2},0)$ & $(\pi,0)$ 
& $(\frac{\pi}{2},\frac{\pi}{2})$ &
$(\pi,\frac{\pi}{2})$ & $(\pi,\pi)$ \\
\hline
$\tilde{n}(\bbox{k})$ & 0.0000 & 0.0586 & 0.7649
& 0.0000 & 0.2269 & 0.0000\\
$n_h(\bbox{k}),A$ & 0.0164  & 0.0237 & 0.2538 
& 0.0658 & 0.2212 & 0.2239\\
$n_h(\bbox{k}),E$ & 0.0069  & 0.0319 & 0.1752 
& 0.0660 & 0.2529 & 0.2675
\end{tabular}
\label{nk}
\end{table}
\begin{table}
\caption{The quantity $|\langle \Psi_c^{(2h)}
|\Psi_0^{(2h)}\rangle|^2$
for all momenta in the $4\times 4$-cluster.}
\begin{tabular}{c | c c c c c c}
$\bbox{k}$ &$(0,0)$ & $(\frac{\pi}{2},0)$ & $(\pi,0)$ 
& $(\frac{\pi}{2},\frac{\pi}{2})$ &
$(\frac{\pi}{2},\pi)$ & $(\pi,\pi)$ \\
\hline
$n(\bbox{k})$ & 0.9846  & 0.9702 & 0.9497 & 0.9445 & 0.9455 & 0.7921
\end{tabular}
\label{antico}
\end{table}
                                                        
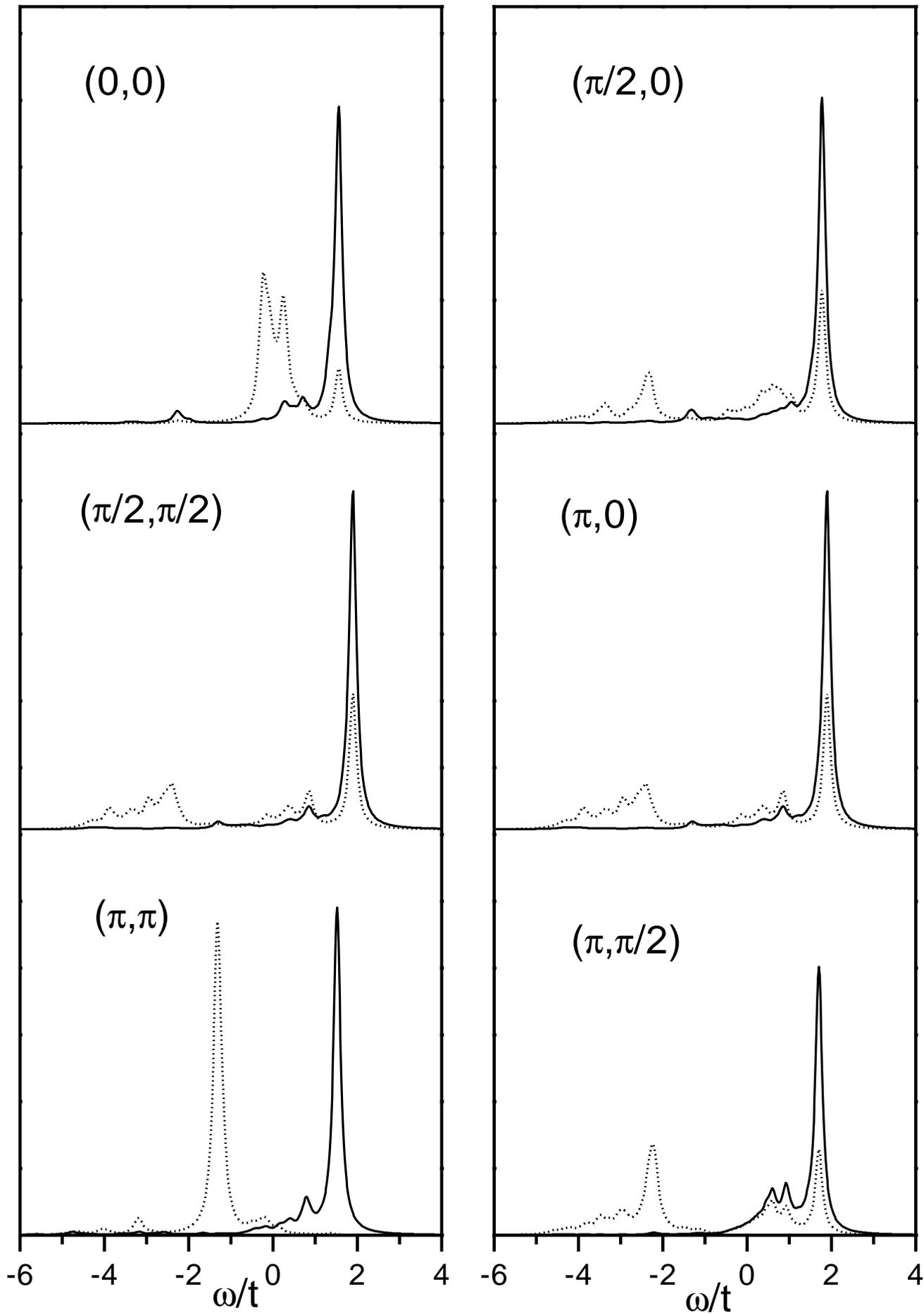
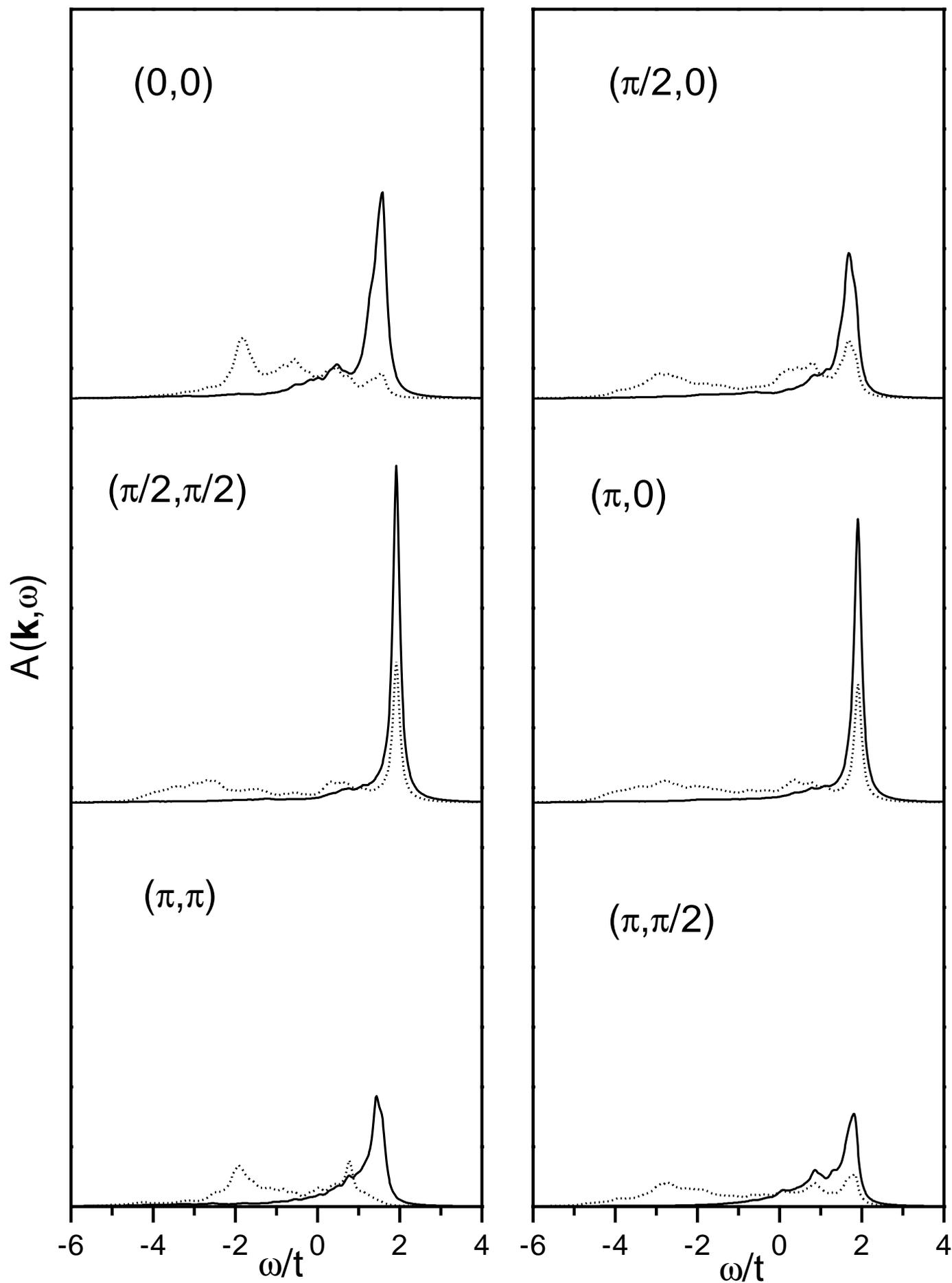
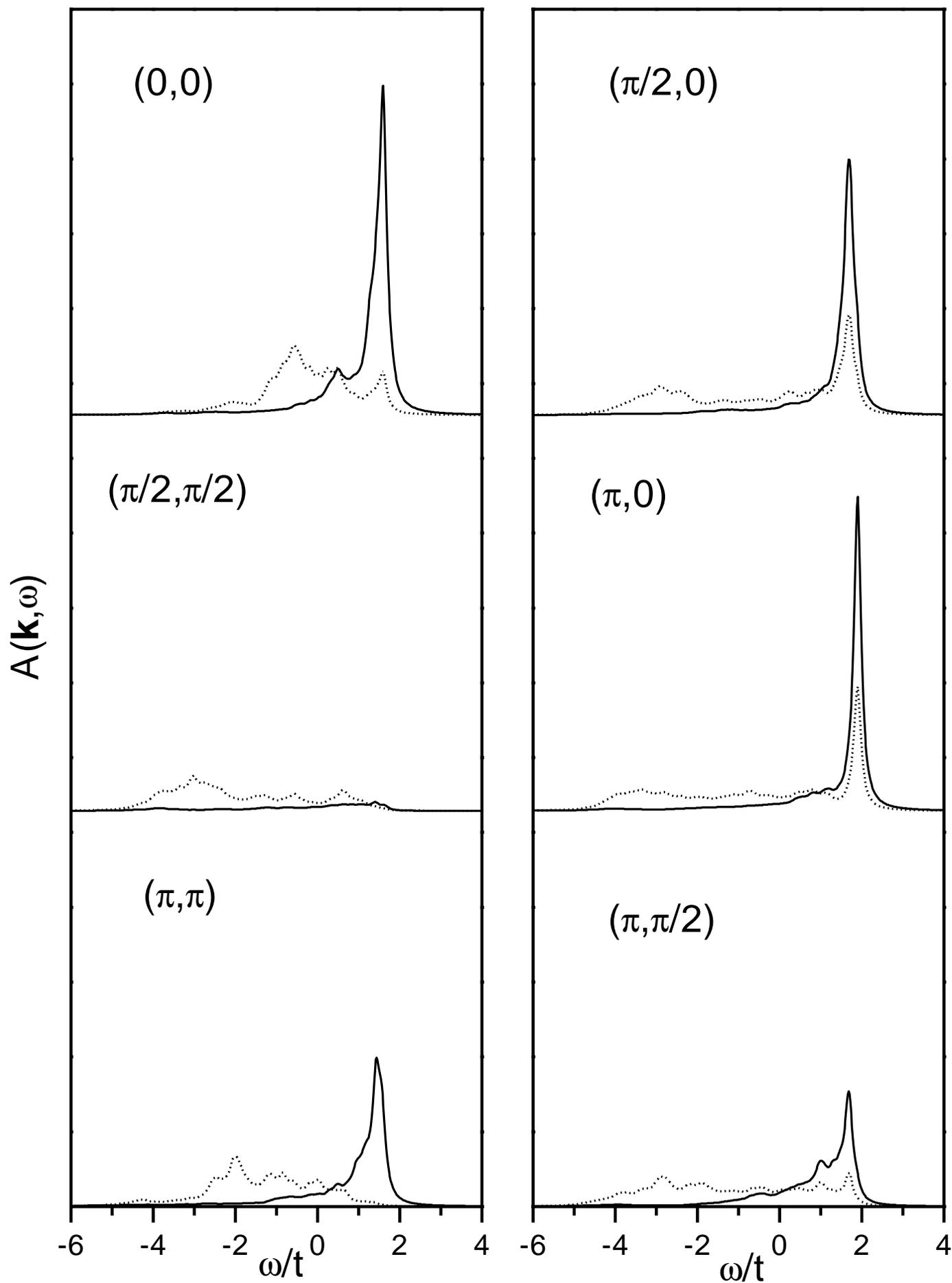
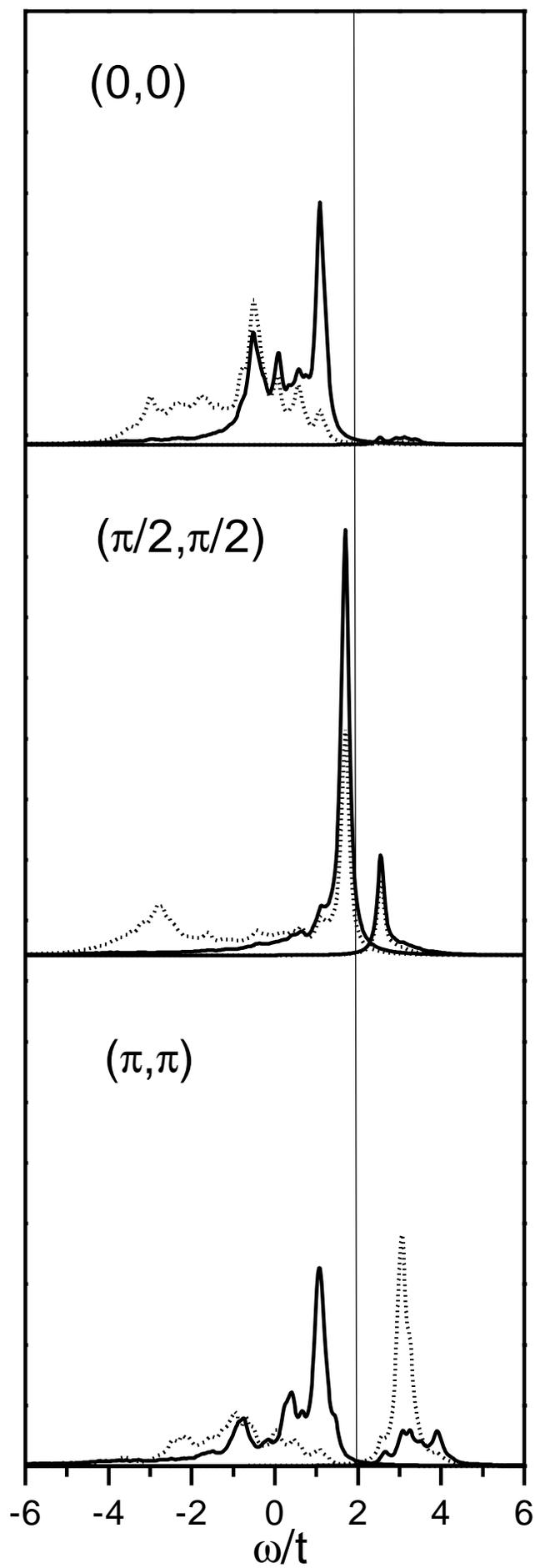
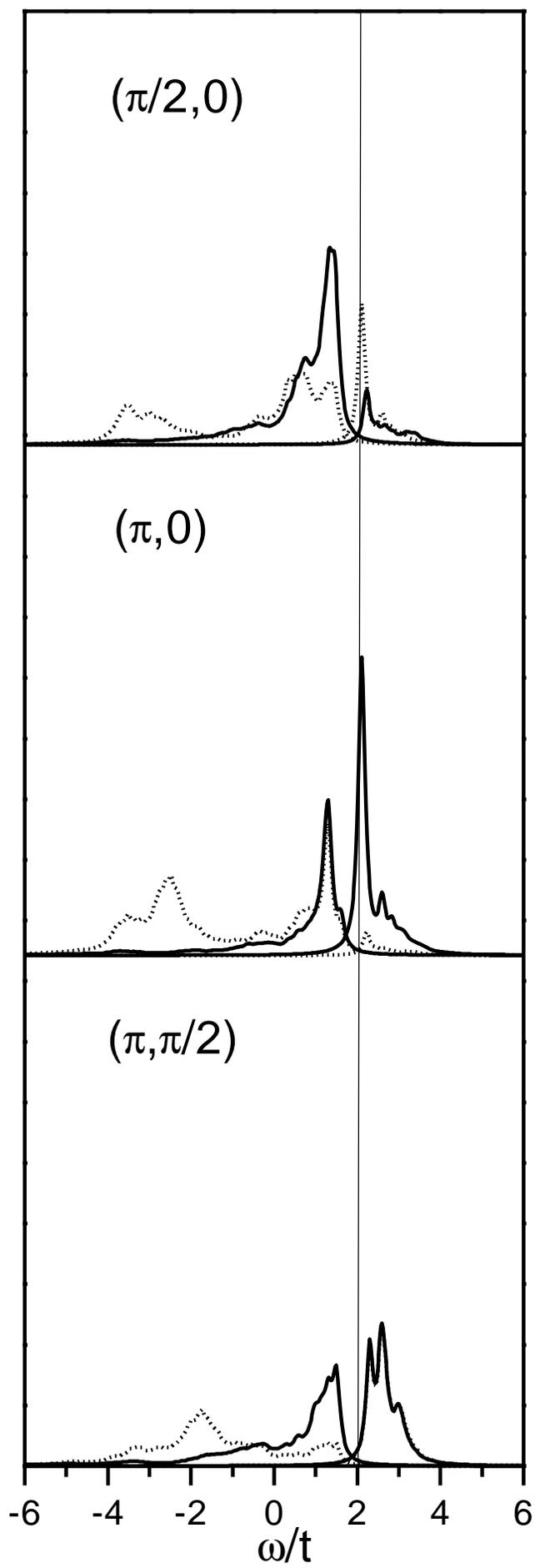
\end{document}